\documentclass[12pt]{article}
\usepackage[T1]{fontenc}
\usepackage{geometry}
\makeatletter

\providecommand{\LyX}{L\kern-.1667em\lower.25em\hbox{Y}\kern-.125emX\@}

\newcommand{\lyxaddress}[1]{
  \par {\raggedright #1 
  \vspace{1.4em}
  \noindent\par}
}

\makeatother

\begin{document}

\title{Defining relations for W-algebras from singular vectors}

\author{Zoltán Bajnok}

\maketitle

\lyxaddress{\emph{Institute for Theoretical Physics, Roland Eötvös University, 1117 Budapest,
Pázmány sétány 1/A, Hungary}}

\begin{abstract}
It is shown that the commutation relations of W-algebras can be recovered from
the singular vectors of their simplest nontrivial, completely degenerate highest
weight representation. 
\end{abstract}

\section{Introduction}

In the paper \cite{Ho} Hornfeck analysed the correspondence between the structure
constants of a W-algebra and the weights of its distinguished completely degenerate
highest weight (h.w.) representation: He supposed the existence of a W-primary
field, \( \Phi  \), such that the operator product expansion of \( \Phi  \)
and any generator of the W-algebra, \( W^{i} \), has the following form:
\begin{equation}
\label{horn}
W^{i}\star \Phi =P_{i}(\partial ^{l}\Phi ,\partial ^{k}W^{j}).
\end{equation}
where on the r.h.s.. only normal ordered differential polynomials of \( \Phi  \)
and the W-generators appear. Analysing the fulfilment of the Jacobi identity
containing the W-modes and the modes of \( \Phi  \) he could determine the
simplest nontrivial structure constants of the W-algebra. The extension of the
procedure for larger class of the structure constants implies that in the consideration
the modes of composite fields have to be included, which makes the computation
involved. Moreover, no general theorem guarantees that the method works, i.e..
the solution exists.  

In our previous paper, \cite{wcor}, we analysed the relation between the classical
and quantum Toda models. We conjectured that at the quantum level -- similarly
to classical case -- the structure of the highest weight representation of the
W-algebra, corresponding to the quantization of the Toda field, determines the
algebra itself. We called this representation the quantum defining representation
since on one hand it is the representation obtained by carrying out the WZW\( \to  \)Toda
reduction procedure in the defining representation of the affine algebra, see
\cite{BaFeFo}. On the other hand its fusion rule reminds of the fusion rule
of the defining representation of the underlying Lie algebra. In the terminology
of Nahm \cite{Na} this is the representation which has the smallest dimensional
nontrivial special subspace (the dimension is equal to the dimension of the
defining representation of the underlying Lie algebra) and the conjecture states
that from the structure of this special subspace the defining relations of the
W-algebra can be recovered. 

In this paper, using the framework of vertex operator algebras, we prove that
the structure of the representation mentioned above, which also has the property
eq. (\ref{horn}) required by Hornfeck determines the algebra itself. 

First we summarize the definitions and properties of vertex operator algebras
tailormade for the applications, then we prove our statement using a lemma,
which is placed in the Appendix. We give examples, which contain among others
the Virasoro and the \( WA_{2} \) algebra and finally conclude.

\section{Vertex operator algebras }

The W-algebras we are dealing with are essentially the WG-algebras, i.e. we
suppose that the generating fields \( W^{j}(z) \) have integer spins as: \( j=2<n_{1}\leq \dots \leq N \).
In describing these W-algebras we use the framework of vertex operator algebras.
A brief summary of their properties are given in order to fix the notation and
collect the results needed. For a detailed description see \cite{FeFr,God}
and references therein. 

A vertex operator algebra is an associative algebra generated by the modes of
the W-fields: \( W^{j}(z)=:\sum _{n}z^{-n-j}W^{j}_{n} \). Usually it is defined
by means of its vacuum module. This module is obtained from the vacuum vector
\( \vert 0\rangle  \) by successively applying the modes \( W^{j}_{n} \).
Each mode acts as a linear operator and

\newcommand{\vac}{\vert 0\rangle }
 
\[
W^{j}_{n}\vac =0\qquad n\geq -j+1.\]
holds. The modes of \( W^{2}(z)=:L(z) \) are distinguished: \( L_{-1} \) generates
translation, \( e^{zL_{-1}}W_{-j}^{j}\vac =W^{j}(z)\vac  \), consequently it
acts as: \def\ad{{\rm ad}}
\begin{equation}
\label{lm1}
L_{-1}W^{j}_{-k}\vac =(k-j+1)W^{j}_{-k-1}\vac \qquad ;\qquad [L_{-1},W^{j}_{-k}]=(k-j+1)W^{j}_{-k-1}.
\end{equation}
 \( L_{0} \) corresponds to the Hamiltonian so it acts diagonally and defines
the grading:
\begin{equation}
\label{l0}
L_{0}W^{j}_{-k}\vac =kW^{j}_{-k}\vac \qquad ;\qquad [L_{0},W^{j}_{-k}]=kW^{j}_{-k}.
\end{equation}
 \( L_{1} \) is the adjoint of \( L_{-1} \) and its action is given by:

\begin{equation}
\label{lp1}
L_{1}W^{j}_{-k}\vac =(j+k-1)W^{j}_{-k+1}\vac \qquad ;\qquad [L_{1},W^{j}_{-k}]=(j+k-1)W^{j}_{-k+1}.
\end{equation}
These modes generate a subalgebra, which is isomorphic to \( sl_{2} \):
\begin{equation}
\label{sl2}
[L_{0},L_{\pm }]=\mp L_{\pm }\quad ;\qquad [L_{1},L_{-1}]=2L_{0}.
\end{equation}
 For the adjoint action \( W^{j}_{j-1} \) is a h.w. vector and \( W^{j}_{-j+1} \)
is a lowest weight (l.w.) vector of a \( 2j-1 \) dimensional representation
of \( sl_{2} \). Any mode from this representation space annihilates the vacuum. 

To each vector of the vacuum module,
\begin{equation}
\label{genvec}
W^{J}_{-K}\vac :=W^{j_{1}}_{-k_{1}}W^{j_{2}}_{-k_{2}}\dots W^{j_{l}}_{-k_{l}}\vac \quad ;\qquad j_{i}\geq j_{i+1}\: ,\; k_{i}\geq k_{i+1}\quad {\rm if}\; j_{i}=j_{i+1}
\end{equation}
a field operator - and element of the vertex operator algebra - \( \Phi ^{h}(z)=\sum _{n}z^{-n-h}\Phi _{n}^{h}\; ,\; h=\sum _{i}k_{i}=\vert K\vert  \),
is associated by the following rule:
\[
e^{zL_{-1}}W^{J}_{-K}\vac =\Phi ^{h}(z)\vac .\]
This operator can be written as a nested normal ordered product of the generating
fields and their derivates as:
\[
\Phi ^{h}(z)=:\frac{\partial ^{n_{1}}}{n_{1}!}W^{j_{1}}(z):\frac{\partial ^{n_{2}}}{n_{2}!}W^{j_{2}}(z):\dots :\frac{\partial ^{n_{i-1}}}{n_{i-1}!}W^{j_{i-1}}(z)\frac{\partial ^{n_{i}}}{n_{i}!}W^{j_{i}}::.\]
 where \( n_{i}=k_{i}-j_{i} \) and the normal ordering is defined as usual:
\[
:\Phi ^{1}(z)\Phi ^{2}(z):=\sum _{n}z^{-n-h_{1}-h_{2}}(:\Phi ^{1}\Phi ^{2}:)_{n}\]
and 
\[
(:\Phi ^{1}\Phi ^{2}:)_{n}=\sum _{k\leq -h_{2}}\Phi ^{2}_{k}\Phi ^{1}_{-k+n}+\sum _{k>-h_{2}}\Phi ^{1}_{-k+n}\Phi ^{2}_{k}.\]

While the relations (\ref{lm1},\ref{l0}) extend for the field \( \Phi ^{h}(z) \)
by the substitution \( W\leftrightarrow \Phi ,\; h\leftrightarrow j \), the
relation (\ref{lp1}) generally does not. (If it is also satisfied by the modes
of a field, then the field is called quasi-primary). The modes \( \Phi _{k}^{h}\: ,\; k=-h+1,\dots ,h-1 \)
all annihilate the vacuum. They form a subalgebra of the vertex operator algebra,
which is denoted by \( W_{s} \). The negative modes which do not annihilate
the vacuum \( \Phi _{k}^{h}\: ,\; k\leq -h \) constitute another subalgebra
of the vertex operator algebra: \( W_{--} \).

The commutation relations of the modes of the fields are encoded in the singular
terms of their operator product expansion:

\def\O{{\cal O}}
\begin{equation}
\label{ope}
\Phi ^{1}(z)\Phi ^{2}(w)=\sum _{k=0}^{h_{1}+h_{2}-1}\frac{\O _{-k}^{12}(w)}{(z-w)^{h_{1}+h_{2}-k}}+O(1)
\end{equation}
 where \( \O ^{12}_{-k}(z)\vac =e^{zL_{-1}}\Phi ^{1}_{h_{1}-k}\Phi ^{2}_{-h_{2}}\vac  \).
Explicitly we have:
\begin{equation}
\label{comrel}
[\Phi ^{1}_{n},\Phi ^{2}_{m}]=\sum _{k=0}^{h_{1}+h_{2}-1}{h_{1}+n-1\choose k}(\O ^{12}_{-h_{1}-h_{2}+k+1})_{n+m}
\end{equation}
It is important to observe that the operator product expansion of quasi-primary
fields, \( \Phi ^{1}(z)\Phi ^{2}(z) \), - and in this way the commutation relations
of their modes - can be recovered from \( \Phi ^{1}_{-h_{1}+1}\Phi ^{2}_{-h_{2}}\vac  \)
and \( \Phi ^{1}_{h_{1}}\Phi ^{2}_{-h_{2}}\vac  \). This can be achieved inductively
by acting with \( L_{1} \) on \( \Phi ^{1}_{-h_{1}+1}\Phi ^{2}_{-h_{2}}\vac  \).
For \( k<2h_{1}-1 \) the result is:
\begin{equation}
\label{lwrel}
L_{1}^{k}\Phi ^{1}_{-h_{1}+1}\Phi ^{2}_{-h_{2}}\vac =\{\prod _{l=1}^{k}(2h_{1}-l)\}\Phi ^{1}_{-h_{1}+k+1}\Phi ^{2}_{-h_{2}}\vac .
\end{equation}
Similarly by acting successively with \( L_{1} \) on \( \Phi ^{1}_{h_{1}}\Phi ^{2}_{-h_{2}}\vac  \)
we have:
\begin{equation}
\label{hwrel}
L_{1}^{k}\Phi ^{1}_{h_{1}}\Phi ^{2}_{-h_{2}}\vac =\{\prod _{l=1}^{k}(-l)\}\Phi ^{1}_{-h_{1}+k+1}\Phi ^{2}_{-h_{2}}\vac 
\end{equation}

\newcommand{\hw}{\vert h.w.\rangle }

A h.w. module for a vertex operator algebra is generated by acting successively
with the modes \( W^{j}_{k} \) on the h.w. vector, \( \hw  \), using the commutation
relations (\ref{comrel}) and the following rules:
\begin{equation}
\label{hwrep}
W^{j}_{k}\hw =0\qquad {\rm for}\quad k>0\quad ;\qquad W^{j}_{0}\hw =w_{j}\hw .
\end{equation}
We adopt the following ordering: the modes of the subalgebra \( W_{s} \) are
on the right of the modes of \( W_{--} \). In both subalgebra we use the same
ordering rule as we used in the vacuum module, (\ref{genvec}). This module
may have singular vectors, vectors which are nontrivial elements of the module
sharing the property (\ref{hwrep}) of the h.w. vector \( \hw  \). Since the
norm of these vectors and their descendants are zero they are set to zero. This
step also ensures the irreducibility of the representation of the vertex operator
algebra. The factor space \( W_{s}\hw /W_{--}\hw  \) is of special interest,
see Nahm \cite{Na} for details. For the vacuum module this factor space is
very simple it is one dimensional.  For the next simplest case, where this factor
space has the smallest nontrivial dimension we formulate our statement.

\section{Main result}

Now the stage is set and we are ready to formulate our statement. The requirement
demanded by Hornfeck ,(\ref{horn}), which is also satisfied by the quantum
Toda representation, can be formulated as the factor space \( W_{s}\hw /W_{--}\hw  \)
is finite dimensional and has basis of the form: \( L_{-1}^{k}\hw \: ,\; k=0,1,\dots ,N-1 \).
In more detail we suppose that:

\begin{equation}
\label{prop}
W^{j}_{-k}\hw =(\sum _{\{J,K\}}\alpha ^{j,k}_{J,K}W^{J}_{-K})\hw \quad ;\qquad {\rm for\, all}\quad 1\leq k<j\: ;\quad 2<j\leq N
\end{equation}
where in \( J=\{j_{1},\dots ,j_{k}\} \) we allow \( j_{i}=1 \) by introducing
\( W^{1}_{-1}=L_{-1} \), in order to unify the notation and shorthen the formula.
The sum is over all \( J \) and \( K \) for which \( 1\leq j_{i}\leq k_{i} \)
and \( \vert K\vert =\sum _{i=1}^{l}k_{i}=k \) hold. We also suppose that 
\begin{equation}
\label{Toda}
W^{N}_{-N}\hw =(\sum _{\{J,K\}}\alpha ^{N,N}_{J,K}W_{-K}^{J})\hw 
\end{equation}
holds, which means that the factorspace is \( N \) dimensional. ( Let us note
that these relations are consequences of the singular vectors). Now we claim
that after taking some assumptions on the coefficients , \( \alpha  \), depending
on the type of the W-algebra, (or equivalently, on the spins of the fields)
the commutation relations of the generating fields, \( [W^{j_{1}}_{k_{1}},W^{j_{2}}_{k_{2}}] \),
can be reproduced. 

As a first step we show that the basis of this module have the form:
\begin{equation}
\label{hwbas}
W^{J}_{-K}L_{-1}^{k}\hw 
\end{equation}
where \( W^{J}_{-K}\in W_{--} \) and \( k<N \). We proceed at each level from
grade to grade. The level of \( W^{J}_{-K} \) is \( \vert K\vert  \) as usual
and its grade is defined to be \( g(W^{J}_{-K})=\vert K\vert -\vert J\vert  \).
At the highest grades we have vectors with \( J=\{j_{1}\} \), which is of the
form (\ref{hwbas}) so the induction starts. A typical element of a h.w. module
has the form \( W^{J}_{-K}W^{L}_{-M}\hw  \), where \( W^{J}_{-K}\in W_{--} \)
and \( W^{L}_{-M}\in W_{s} \). When we transform this vector into the basis
(\ref{hwbas}) we use the following steps: either we commute the operators into
the right order, or we use the relation (\ref{prop}) in order to replace the
modes of \( W_{s} \) with the modes of \( W_{--} \) and with \( L_{-1}^{n} \),
or we use the relation (\ref{Toda}) in order to replace \( L_{-1}^{n} \) for
\( n>N \) with the modes of \( W_{--} \) and with \( L_{-1}^{k} \), where
now \( k<N \). From the relations (\ref{ope},\ref{comrel}) it follows, that
the grade of any of the monoms appearing in the commutator of two modes is always
greater than the sum of the grades. Moreover this also holds for the other steps,
that is the grade is always increasing. Consequently the basis can be reached
inductively. 

In proving the statement we will determine \( [W^{j_{1}}_{k_{1}},W^{j_{2}}_{k_{2}}] \)
by induction in \( G=j_{1}+j_{2}\: ;\; j_{1}<j_{2} \) and for fixed \( G \)
in \( j_{1} \). We suppose that all the commutators \( [W^{j_{1}}_{k_{1}},W^{j_{2}}_{k_{2}}] \)
are known for any \( j_{1}+j_{2}<G \) and for \( G=j_{1}+j_{2} \) for any
\( j_{1}<j \) and by means of this hypothesis we proceed either for \( G \)
and \( j_{1}=j \) or for \( G+1=2+(G-1) \). 

Since the generating fields are quasi primary fields it is enough to determine
\( (\O ^{j_{1}j_{2}}_{-j_{1}-j_{2}+1})_{-j_{1}-j_{2}+1}\vac =W^{j_{1}}_{-j_{1}+1}W^{j_{2}}_{-j_{2}}\vac  \)
and \( (\O ^{j_{1}j_{2}}_{j_{1}-j_{2}})_{j_{1}-j_{2}}\vac =W^{j_{1}}_{j_{1}}W^{j_{2}}_{-j_{2}}\vac  \). 

We start with \( \O ^{j_{1}j_{2}}_{-j_{1}-j_{2}+1} \). Consider the equation:
\begin{equation}
\label{pr-com}
(\O ^{j_{1}j_{2}}_{-j_{1}-j_{2}+1})_{-j_{1}-j_{2}+1+k}\hw =[W^{j_{1}}_{-j_{1}+1},W^{j_{2}}_{-j_{2}+k}]\hw \: ,\; k=1,\dots ,j_{1}+j_{2}-1
\end{equation}
which is a consequence of (\ref{comrel}). On the r.h.s. of (\ref{pr-com})
we replace the commutator with the difference of the products and in each term
we use the relation (\ref{prop}) in order to replace \( W^{j_{2}}_{-j_{2}+k} \),
(\( W^{j_{1}}_{-j_{1}+1} \)) with the modes of \( W_{--} \) and with \( L_{-1}^{k} \)
for \( k<N \). After the replacement the spin of the operators appearing is
less then \( j_{2} \) , (\( j_{1} \) ), consequently the induction hypothesis
can be used to transform the product into the basis (\ref{hwbas}). Now we need
a lemma. We prove in the appendix, that \( \Phi ^{h}_{-h+1}\hw \: ,\; \Phi ^{h}_{-h+2}\hw \: ,\; \dots \: ,\; \Phi ^{h}_{0}\hw  \)
completely determines \( \Phi ^{h}(z) \). Applying the lemma for the present
case the operator \( \O ^{j_{1}j_{2}}_{-j_{1}-j_{2}+1}(z) \) is completely
determined. By means of equation (\ref{lwrel}) the operators \( \O ^{j_{1}j_{2}}_{-j_{1}-j_{2}+k+1}(z)\: ,\; k=0,\dots ,2j_{1}-2 \)
can be determined as well. 

The next step is to compute \( \O ^{j_{1}j_{2}}_{j_{1}-j_{2}}=W^{j_{1}}_{j_{1}}W^{j_{2}}_{-j_{2}}\vac  \).
The strategy is the same as before. We consider the equation:
\begin{equation}
\label{pr+com}
[W^{j_{1}}_{j_{1}},W^{j_{2}}_{-j_{2}+k}]\hw \sim (\O ^{j_{1}j_{2}}_{j_{1}-j_{2}})_{j_{1}-j_{2}+k}\hw \: ,\; k=1,\dots ,j_{2}-j_{1}
\end{equation}
where by \( \sim  \) we mean that modulo the already known operators \( \O ^{j_{1}j_{2}}_{-j_{1}-j_{2}+k+1}(z)\: ,\; k=0,\dots ,2j_{1}-2 \).
We proceed the same way and replace the commutator with the difference of the
products and use the relation (\ref{prop}). This second step is slightly different
from the previous one. Clearly for \( j_{1}=j_{2} \) (in the last induction
step for fixed \( G \)) we need to compute \( [W^{j_{1}}_{j_{1}},W^{j_{1}}_{-j_{1}}] \)
in order to determine \( \O ^{j_{1}j_{1}}_{0} \). The problem is that we cannot
use the lemma for this case. Note however, that the central term,\( \O ^{j_{1}j_{2}}_{0} \),
in the OPE of \( W^{j}(z)W^{j}(w) \) contributes only in \( [W^{j_{1}}_{j_{1}+k},W^{j_{1}}_{-j_{1}-k}] \).
This makes it possible to go further in the induction for the next \( j_{1}^{'}+j_{2}^{'}=G^{'}>G=2j_{1} \).
In computing \( [W^{j^{'}_{1}}_{-j^{'}_{1}},W^{j^{'}_{2}}_{-j^{'}_{2}+k}] \)
and \( [W^{j^{'}_{1}}_{j^{'}_{1}},W^{j^{'}_{2}}_{-j^{'}_{2}+k}] \) for \( j_{1}^{'}<j_{1} \)
the same way as before the central term \( \O ^{j_{1}j_{1}}_{0} \) can contribute
via terms like \( (W^{k}_{-l}W^{j_{1}}_{j_{1}+n})W^{j_{1}}_{-j_{1}-n} \) contained
in the composite operator \( :W^{k}(z)W^{j_{1}}(z): \). Since the composite
operators appear in commutators, (where the sum of the spins is always decreasing)
and \( k \) is at least \( 2 \) they do not contribute if \( G^{'}\leq G+2 \).
We restrict ourselves to this case. This covers the WG-algebras from the infinite
series. The exceptional cases has to be considered separately. We also make
a difference depending on whether \( G^{'}=G+1 \) or \( G^{'}=G+2 \). In the
first case we consider eq. (\ref{pr+com}) for \( j_{1} \) and \( j_{2}=j_{1}+1 \)
with \( k=1 \). Since \( \O ^{j_{1}j_{2}}_{-1}=0 \) we can recover \( \O ^{j_{1}j_{1}}_{0} \)
if \( \alpha ^{j_{1}+1,j_{1}}_{j_{1},j_{1}} \) is not zero. In the second case
we have to consider eq. (\ref{pr+com}) for \( j_{1} \) and \( j_{2}=j_{1}+2 \)
with \( k=1,2 \). We demand from the coefficients, \( \alpha  \), that these
two equation can be solved for the two unknowns: \( \O _{-2}^{j_{1}j_{1}+2} \)
and \( \O _{-2}^{j_{1}j_{1}+2} \). ( Similar but more complicated conditions
has to be demanded in the exceptional cases). See also the example of \( \O ^{24}_{-2} \)
in the next section. Having determined \( \O ^{j_{1}j_{2}}_{j_{1}-j_{2}} \)
we use relation (\ref{hwrel}) to obtain \( \O ^{j_{1}j_{2}}_{j_{1}-j_{2}+k} \),
with which the commutator \( [W^{j_{1}}_{i_{1}},W^{j_{2}}_{i_{2}}] \) is determined
and the theorem is proved.

\section{Examples}

The induction procedure starts with \( G=2+2 \) , that is with the commutation
relations of the modes of \( L(z) \). Since this is the basic field of the
vertex operator algebra in computing its commutation relation it is not necessary
to use the method developed. However we follow the general scheme in order to
see easier how to proceed in the general case. We start to compute \( \O ^{22}_{-3}(z) \)
or equivalently to compute \( (\O ^{22}_{-3})_{-3}\vac  \). At level three
we have two operators: \( W^{3}_{-3}\vac  \) and \( L_{-1}L_{-2}\vac  \),
written in the quasi primary basis. From the relation 
\[
[L_{-1},L_{-1}]\hw =(\O ^{22}_{-3})_{-2}\hw =0\]
 we realize that there is no \( W^{3}_{-3}\vac  \) term. From 
\[
[L_{-1},L_{0}]\hw =(\O ^{22}_{-3})_{-1}\hw =-L_{-1}\hw \]
 we can fix \( (\O ^{22}_{-3})_{-3}\vac  \) to be \( L_{-1}L_{-2}\vac  \),
which is also consistent with 
\[
[L_{-1},L_{1}]\hw =(\O ^{22}_{-3})_{0}\hw =-2w_{2}\hw .\]
Now acting with \( L_{1} \) on the relation 
\[
(\O ^{22}_{-3})_{-3}\vac =L_{-1}L_{-2}\vac \]
we have 
\[
(\O ^{22}_{-2})_{-2}\vac =2L_{-2}\vac \]
and acting once more we have 
\[
(\O ^{22}_{-1})_{-1}\vac =0.\]
The next step is to compute \( (\O ^{22}_{0})_{0} \). As we explained in the
general case this needs special care. We discuss the individual cases separately.
If we are analysing the Virasoro algebra itself, then the relation (\ref{Toda})
takes the following form:
\[
L_{-2}\hw =\alpha ^{22}_{\{(2,1),(2,1)\}}L_{-1}L_{-1}\hw \]
Applying \( L_{2} \) on both sides of the relation we can obtain:
\[
(\O ^{22}_{0})_{0}=6w_{2}\alpha ^{22}_{\{(1,1),(1,1)\}}-4w_{2}\]
Substituting the usual parameterization, \( \alpha ^{22}_{\{(2,1),(2,1)\}}=t \)
and \( w_{2}=\frac{3}{4t}-\frac{1}{2} \) we recover \( c=13-6t-t^{-1} \). 

If we are analysing some other W-algebra we distinguish whether the next spin
is \( 3 \) or \( 4 \), i.e. \( G=G+1 \) or \( G=G+2 \) referring to the
general case. In the first case we go on in the induction and compute \( (\O ^{23}_{-4})_{-4}\vac  \).
The operators in the quasi primary basis at level four correspond to the action
of the following modes on the vacuum vector: \( W^{4}_{-4}\, ,\: L_{-2}L_{-2}-\frac{3}{10}L_{-1}^{2}L_{-2}\, ,\: L_{-1}W^{3}_{-3}\, ,\: L_{-1}^{2}L_{-2} \).
From 
\[
[L_{-1},W^{3}_{-2}]=(\O ^{23}_{-4})_{-3}=0\]
it follows that we do not have any non-derivative field. The relation
\[
[L_{-1},W^{3}_{-1}]=(\O ^{23}_{-4})_{-2}=-W^{3}_{-2}\]
together with 
\[
[L_{-1},W^{3}_{0}]=(\O ^{23}_{-4})_{-1}=-2W^{3}_{-1}\quad ;\quad [L_{-1},W^{3}_{1}]=(\O ^{23}_{-4})_{0}=-3W^{3}_{0}\]
give \( (\O ^{23}_{-4})_{-4}\vac =L_{-1}W^{3}_{-3}\vac  \). From this we can
recover \( (\O ^{23}_{-3})_{-3}\vac =3W^{3}_{-3}\vac  \) and \( (\O ^{23}_{-2})_{-2}\vac =0 \),
by applying \( L_{1} \). The next step is to compute \( L_{2}W^{3}_{-3}\vac =(\O ^{23}_{-1})_{-1}\vac  \).
Since there is no operator at level one this term is zero and the commutation
relations of the modes of \( L(z) \) and \( W^{3}(z) \) are completely determined:
\begin{equation}
\label{lwcom}
[L_{n},W^{3}_{m}]=(2n-m)W^{3}_{n+m}.
\end{equation}
 Now \( L_{2} \) can be applied to the relation 
\begin{equation}
\label{w32}
W^{3}_{-2}\hw =(\alpha ^{32}_{\{(2,2)\}}L_{-2}+\alpha ^{32}_{\{(1,1),(1,1)\}}L_{-1}L_{-1})\hw 
\end{equation}
in order to compute 
\begin{equation}
\label{wcent}
\O ^{22}_{0}=(\alpha _{\{(2,2)\}}^{32})^{-1}(6w_{3}-6\alpha ^{32}_{\{(1,1),(1,1)\}}w_{2})-4w_{2}.
\end{equation}

Either we are considering the \( G=G+2 \) case or some other W-algebra with
spins \( 2,3,4,\dots  \) the next step is to compute the commutation relations
of the modes of \( L(z) \) and \( W^{4}(z) \). The determination of \( (O^{24}_{-5})_{-5}\vac  \)
goes as before and gives \( L_{-1}W^{4}_{-4}\vac  \). In order to determine
\( (\O ^{24}_{-2})_{-2}\vac =\beta L_{-2}\vac  \) we consider the relation
\[
[L_{2},W^{4}_{-3}]=(\O ^{24}_{-5})_{-1}+3(\O ^{24}_{-4})_{-1}+3(\O ^{24}_{-3})_{-1}+(\O ^{24}_{-2})_{-1}=9W^{4}_{-1}+\beta L_{-1}\]
and the relation 
\[
[L_{2},W^{4}_{-2}]=(\O ^{24}_{-5})_{0}+3(\O ^{24}_{-4})_{0}+3(\O ^{24}_{-3})_{0}+(\O ^{24}_{-2})_{0}=8W^{4}_{0}+\beta L_{0}\]
simultaneously. We apply both commutators on the vector \( \hw  \) and apply
the procedure introduced above. Denoting the central term \( O^{22}_{0} \)
by \( \gamma  \) we have the following equations:
\[
\beta w_{2}-\gamma \alpha ^{42}_{\{(2,2)\}}=6w_{2}+\alpha ^{42}_{\{(2,2)\}}-8w_{4}\]
and 
\[
\beta -\gamma \alpha ^{43}_{\{(2,2),(2,1)\}}=\alpha ^{31}_{\{(2,1)\}}(-9+7\alpha ^{43}_{\{(3,3)\}})+5\alpha ^{43}_{\{(2,3)\}}+4(w_{2}+1)\alpha ^{43}_{\{(2,2),(2,1)\}}+6(3w_{2}+1)\alpha ^{42}_{\{(2,1),(2,1),(2,1)\}}\]
If we have a W-algebra of type \( 2,4,\dots  \) then \( \alpha ^{43}_{\{(3,3)\}} \)is
zero and we demand from the coefficients that the these equations be soluble
for \( \beta  \) and \( \gamma  \). If we are analysing a W-algebra of type
\( 2,3,4,\dots  \) then \( \gamma  \) is already known, (\ref{wcent}), and
any of the equations above can be used to determine \( \beta  \), the other
gives relation among the coefficients. 

Let us finish this part by computing the commutation relations \( [W^{3}_{n},W^{3}_{m}] \)
which is the next one in the induction. Note also that this is the first place
where the application of the general procedure is not formal. The relations
(\ref{prop}) in our case are (\ref{w32}) and : 
\begin{equation}
\label{w331}
W^{3}_{-1}\hw =\alpha ^{31}_{\{(2,1)\}}L_{-1}\hw 
\end{equation}
Using these relations we need to compute \( (\O ^{33}_{-5})_{-5}\vac  \). The
operators at level five corresponds to the following modes: \( W^{5}_{-5}\, ,\: W^{3}_{-3}L_{-2\, ,\: }L_{-1}W^{4}_{-4\, ,\: }L_{-1}L_{-2}L_{-2}\, ,\: L_{-1}^{2}W^{3}_{-3}\, ,\: L_{-1}^{3}L_{-2} \).
From the commutator 
\[
[W^{3}_{-2},W^{3}_{-2}]=(\O ^{33}_{-5})_{-4}=0\]
 we can see that in the OPE there is no nonderivative field. We compute 
\[
[W_{-2}^{3},W^{3}_{-1}]=(\O ^{33}_{-5})_{-3}=:\delta (L_{-1}W^{4}_{-4})_{-3}+\kappa (L_{-1}L_{-2}L_{-2})_{-3}\]
by replacing the commutator with \( W_{-2}^{3}W^{3}_{-1}-W_{-1}^{3}W^{3}_{-2} \)
and applying it on the vector \( \hw  \). We use relations (\ref{w331},\ref{w32})
in order to replace the \( W^{3}(z) \) modes with the Virasoro modes then we
commute the W-modes through the Virasoro modes and use relations (\ref{w331},\ref{w32})
once more. Finally the vector is written in the basis (\ref{hwbas}) as: 
\begin{equation}
\label{w2w1}
\{\alpha ^{32}_{\{(2,2)\}}(-3W^{3}_{-3}+(\alpha ^{31}_{\{(2,1)\}}-2\alpha ^{32}_{\{(2,1),(2,1)\}})L_{-3}-2\alpha ^{32}_{\{(2,1),(2,1)\}}L_{-2}L_{-1})-2(\alpha ^{32}_{\{(2,1),(2,1)\}})^{2}L_{-1}^{3}\}\hw 
\end{equation}
Now using the inverse of the map \( \varphi  \) defined in the appendix we
obtain \( \delta =3\alpha ^{32}_{\{(2,2)\}}/\alpha ^{43}_{\{(3,3)\}} \) and
\( \kappa =\alpha ^{32}_{\{(2,1),(2,1)\}} \). The computation of 
\[
[W_{-2}^{3},W^{3}_{0}]=(\O ^{33}_{-5})_{-2}=:\delta (L_{-1}W^{4}_{-4})_{-2}+\kappa (L_{-1}L_{-2}L_{-2})_{-2}+\mu (L_{-1}W^{3}_{-3})_{-2}\]
follows the same line as the previous one and gives: \( \mu =2\alpha ^{32}_{\{(2,2)\}}+\alpha ^{32}_{\{(2,1),(2,1)\}}+(\alpha ^{32}_{\{(2,2)\}})^{-1}(\delta \alpha ^{42}_{\{(2,2)\}}+2\kappa h) \).
An analogous way we can compute \( [W_{-2}^{3},W^{3}_{1}]=(\O ^{33}_{-5})_{-1} \)
and determine the coefficient of \( L_{-1}^{3}L_{-2} \), which is
\[
1/6(6\mu \alpha ^{31}_{\{(2,1)\}}-3(2h+1)\kappa -3\alpha ^{41}_{\{(2,1)\}}\delta +6\alpha ^{32}_{\{(2,1),(2,1)\}}w_{3}+\alpha ^{31}_{\{(2,1)\}}(5\alpha ^{32}_{\{(2,2)\}}+6\alpha ^{32}_{\{(2,1),(2,1)\}})\]
Now applying \( L_{1} \) on the vector \( (\O ^{33}_{-5})_{-5}\vac  \) we
can determine \( (\O ^{33}_{-k})_{-k}\vac \quad ;\quad k=1,2,3,4 \). The computation
of \( (\O ^{33}_{0})_{0}\vac  \) is a bit complicated but it is very similar
to the determination of \( (\O ^{22}_{0})_{0}\vac  \). 

Let us specify the results obtained for the \( WA_{2} \) algebra. In this case
we have an extra relation (\ref{Toda}):
\begin{equation}
\label{w3toda}
W^{3}_{-3}\hw =(\alpha ^{33}_{\{(2,3)\}}L_{-3}+\alpha ^{33}_{\{(2,2),(2,1)\}}L_{-2}L_{-1}+\alpha ^{33}_{\{(2,1),(2,1),(2,1)\}}L_{-1}^{3})\hw 
\end{equation}
which has to be used to replace \( L_{-1}^{3} \)with the modes of \( W_{--} \)
in (\ref{w2w1}). The highest weights of this representation can be parametrized
as:
\[
h=\frac{4}{3t^{2}}-1\quad ;\qquad w_{3}=-\frac{(5-3t^{2})(4-3t^{2})}{27t^{3}}\]
Substituting the following values for the coefficients of the singular vectors
\cite{A2}:
\[
\alpha ^{31}_{\{(2,1)\}}=(\frac{t}{2}-\frac{5}{6t})\quad ;\qquad \alpha ^{32}_{\{(2,2)\}}=\frac{2}{3t}\quad ;\qquad \alpha ^{32}_{\{(2,1),(2,1)\}}=-t\]
 
\[
\alpha ^{33}_{\{(2,3)\}}=\frac{1}{6t}+\frac{t}{2}\quad ;\qquad \alpha ^{33}_{\{(2,2),(2,1)\}}=t\quad ;\qquad \alpha ^{33}_{\{(2,1),(2,1),(2,1)\}}=-t^{3}\]
 we obtain \( \delta =0\, ,\: \kappa =\frac{1}{3}\, ,\: \mu =0 \) and the coefficient
of \( L^{3}_{-1}L_{2} \) is \( \frac{1}{6}(\frac{5}{3}-t^{2}-t^{-2}) \). They
give the usual relations of the W-algebra \cite{Zam}, with \( c=2-24(t-t^{-1})^{2} \),
which can be read off from (\ref{wcent}), the only difference is that \( W^{3}(z) \)
is rescaled by the factor of \( \sqrt{\frac{22+5c}{48}} \).

\section{Conclusion}

We have shown in this paper, that the commutation relations of some W-algebras
can be reconstructed from the singular vectors of their special h.w. representation.
We have supposed that the weights of this representation as well as the coefficients
of the singular vectors are given. We have not addressed the question how to
determine them, or whether they give consistent equations. A work is in progress
to answer these questions.

\section*{Appendix}

We would like to prove that \( \Phi ^{h}_{-h+1}\hw ,\Phi ^{h}_{-h+2}\hw ,\dots ,\Phi ^{h}_{0}\hw  \)
completely determines \( \Phi ^{h}_{-h}\vac  \). From now on we suppress the
spin of \( \Phi  \), we write \( \Phi _{-h} \) instead of \( \Phi ^{h}_{-h} \).
First we use the basis related to the action of the \( sl_{2} \) algebra, (\ref{sl2}),
that is we write

\[
\Phi _{-h}\vac =(\Phi _{-h}^{(h)}+\ad L_{-1}\Phi _{-h+1}^{(h-1)}+\dots +\ad ^{k}L_{-1}\Phi ^{(h-k)}_{-h+k}+\dots )\vac \]
where \( \Phi ^{(h-k)}(z) \) is defined such a way that \( \ad L_{1}\Phi ^{(h-k)}(z)=0 \).
The fields can be determined recursively as follows: First note that \( L^{n}_{1}L^{n}_{-1}\Phi _{-h}\vac =c(n,h)\Phi _{-h}\vac  \),
where \( c(n,h)=\prod _{k=1}^{n}k(2h+n-1) \). Now we have:
\[
\Phi _{-2}^{(2)}\vac =c^{-1}(h-2,2)L_{1}^{h-2}\Phi _{-h}\vac \]
 
\[
\Phi _{-3}^{(3)}\vac =c^{-1}(h-3,3)L_{1}^{h-3}(\Phi _{-h}-L_{-1}^{h-2}\Phi _{-2}^{(2)}\vac \]
 and in general 
\[
\Phi _{-k}^{(k)}\vac =c^{-1}(h-k,k)L_{1}^{h-k}(\Phi _{-h}-L_{-1}^{h-k+1}\Phi _{-k+1}^{(k-1)}-\dots -L_{-1}^{h-2}\Phi _{-2}^{(2)})\vac \]
From the definition of the derivative of a field it follows, that
\begin{equation}
\label{negdeg}
(\Phi ^{(h-k)}_{-h}(z))_{-h+l}=0\quad ;\qquad l=1,2,\dots ,k
\end{equation}
This shows that the first place where \( \Phi ^{(h-k)}(z) \) contributes is
at level \( h-k+1 \), consequently we have to determine it from \( \Phi _{-h+k+1}\hw  \).
So we will determine these terms recursively from level to level: \( \Phi ^{(h)}(z) \)
from \( \Phi _{-h+1}\hw  \), \( \Phi ^{(h-1)}(z) \) from \( \Phi _{-h+2}\hw  \),
etc.. To prove this we introduce an ordering. We say that 
\[
W^{j_{1}}_{-i_{1}}\dots W^{j_{k}}_{-i_{k}}\vac \prec W^{l_{1}}_{-m_{1}}\dots W^{l_{p}}_{-m_{p}}\vac \]
if the level is smaller, \( \vert I\vert <\vert M\vert  \) or on the same level
if the grade introduced earlier is smaller, \( \vert I\vert -\vert J\vert <\vert L\vert -\vert M\vert  \),
or if they are on the same level and grade we use the lexicographical ordering,
that is if \( j_{n}>l_{n} \) for the first \( n \) in the row, or if \( j_{n}=l_{n} \)
for all \( n \), then if \( i_{n}<m_{n} \) for the first \( n \) in the row.
Note that we have an analogous ordering in the module \( \hw  \), considering
\( L_{-1} \) to be \( W^{1}_{-1} \). 

Since we are interested in the operators from the factor space \( W_{--}/L_{-1}W_{--} \)
we need the explicit form for its basis. First we fix \( (j_{1},j_{2},\dots ,j_{k-1},j_{k}). \)
We distinguish between two cases: \( j_{k-1}>j_{k} \)and \( j_{k-1}=j_{k} \)
. In the first case at level \( N \) the basis can be written as \( (n_{1},\dots ,n_{k}) \),
where \( \sum _{l}n_{l}=N \) and the \( n \) -s corresponding to same \( j \)
-s are ordered. These vectors are mapped to the level \( N+1 \) space by \( \partial  \).
The kernel of \( \partial  \) is zero since it consists only of the constants.
There is an invertible map from the image of \( \partial  \) to the following
space spanned by the vectors of the form \( (n_{1},\dots ,n_{k}+1) \) explicitly
\( \partial (n_{1},\dots ,n_{k})\mapsto (n_{1},\dots ,n_{k}+1) \). The complement
of this space is spanned by vectors of the form \( (n_{1},\dots ,n_{k-1},0) \),
which gives the basis needed. In the case where \( j_{k-1}=j_{k} \) the basis
at level \( N \) looks like \( (n_{1},\dots ,n_{k-1},n_{k}) \) where \( n_{k-1}\leq n_{k} \)
and \( \sum _{l}n_{l}=N \).We have a map defined the same way as before \( \partial (n_{1},\dots ,n_{k})\mapsto (n_{1},\dots ,n_{k}+1) \),
however the complement looks like \( (n_{1},\dots ,n_{k-1},n_{k-1}) \), and
gives the basis for \( W_{--}/L_{-1}W_{--} \). 

Now we define a map from \( W_{--}/L_{-1}W_{--} \) to the h.w. module by:
\[
v=W^{j_{1}}_{-i_{1}}\dots W^{j_{k}}_{-i_{k}}\vac \mapsto \varphi (v)=\left\{ \begin{array}{c}
W^{j_{1}}_{-i_{1}}\dots W^{j_{k}-1}_{-j_{k}+1}\hw \; if\quad j_{k-1}>j_{k}\\
W^{j_{1}}_{-i_{1}}\dots W_{-2i_{k}+j_{k}}^{j_{k}}W^{j_{k}-1}_{-j_{k}+1}\hw \; if\quad j_{k-1}=j_{k}
\end{array}\right) \]
The properties of the map is that if \( u\prec v \) then \( \varphi (u)\prec \varphi (v) \)
moreover \( \varphi (v) \) is the minimal element of the terms of \( (W^{j_{1}}_{-i_{1}}\dots W^{j_{k}}_{-i_{k}})_{-\sum _{l}i_{l}+1}\hw  \).
It is also true that \( g(\varphi (v))=g(v) \). 

In order to prove this we note that \( min(g([W^{j_{1}}_{i_{1}},W^{j_{2}}_{i_{2}}]))>g(W^{j_{1}}_{i_{1}})+g(W^{j_{2}}_{i_{2}}) \)
so if we are looking for the smaller grade terms we can suppose that the operators
commute. Consequently we do not have to use the normal ordering we can simply
write:

\[
(W^{j_{1}}_{-i_{1}}\dots W^{j_{k}}_{-i_{k}})_{-\sum _{l=1}i_{l}+1}\hw =\sum _{N;\vert N\vert =\vert I\vert -1}c(N)W^{j_{1}}_{-n_{1}}\dots W^{j_{k}}_{-n_{k}}\hw \]
 which is exact modulo higher grade terms. Each term in the sum has the same
degree, which is \( g(W^{j_{1}}_{-i_{1}}\dots W^{j_{k}}_{-i_{k}})-1 \). Clearly
at least for one of the operators \( n_{l}<j_{l} \), so this term has a negative
degree. Using relation (\ref{prop}) terms of this type are replaced with terms
of non-negative degree. This shows that the degree is always increasing by at
least one. It is increasing exactly by one if for only one \( l \) the relation
\( n_{l}=i_{l}-1 \) holds and for all the others \( n_{l}=i_{l} \). This is
a consequence of the fact that each term which contains term with \( j_{l}\leq n_{l}<i_{l} \)
has zero \( c(N) \) coefficient, see (\ref{negdeg}). Now it is easy to see
that \( g(\varphi (v))=g(v) \)and that the minimal element in the ordering
is exactly \( \varphi (v) \), since in all the other cases some higher or never
smaller (\( j_{l}\geq j_{k} \)) spin operator \( (W^{j_{l}}_{-n_{l}})_{-j_{l}+1} \)
is replaced by \( W_{-j_{l}+1}^{j_{l}-1} \). 

The advantage of the ordering defined above is that higher terms give contribution
to higher vectors in the h.w. module. This makes possible to determine them
recursively starting from the lower terms, or with other words the map \( \varphi  \)
can be inverted.

\end{document}